# Drying and deposition of poly(ethylene oxide) droplets determined by Péclet number

Kyle Anthony Baldwin[a], Manon Granjard[b], David Willmer[a],
Khellil Sefiane[b], David John Fairhurst[*a]

[a]Nottingham Trent University, Clifton Lane, Nottingham, NG11 8NS, UK.
[b]University of Edinburgh, Mayfield Road, Edinburgh, EH9 3JL, UK

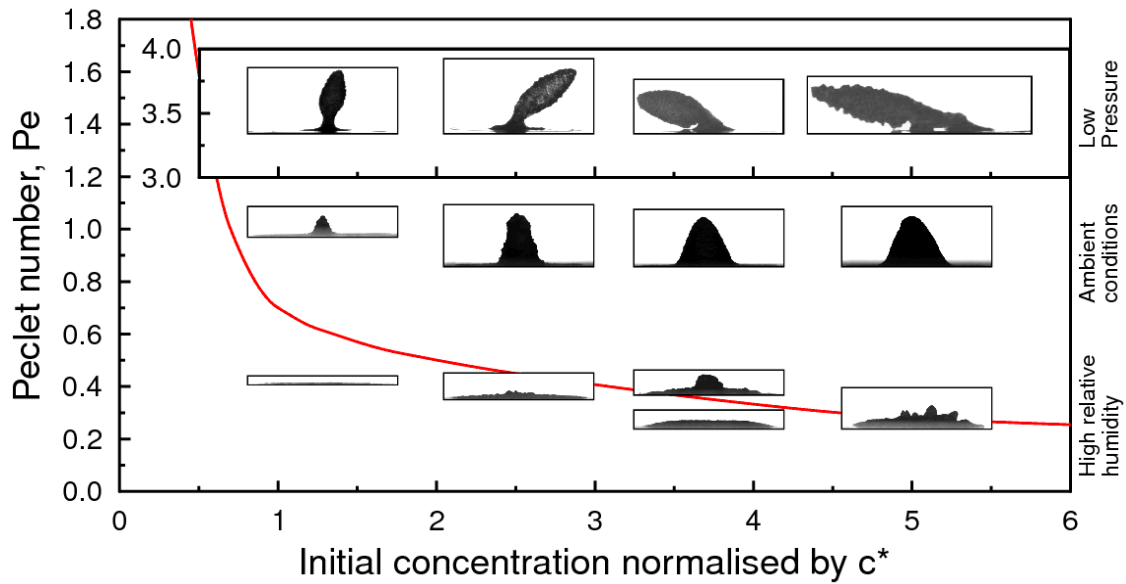

**By varying many experimental parameters (temperature, pressure, humidity, contact angle, concentration and volume) we discover that whether evaporating droplets of PEO polymer solution deposit tall solid pillars or flat puddles is controlled by the dimensionless Péclet number, relating flux to diffusion.**

**Abstract**

We report results of a detailed experimental investigation into the drying of sessile droplets of aqueous poly(ethylene oxide) (PEO) polymer solutions under various experimental conditions. Samples are prepared with a range of initial concentrations $c_0$ and are filtered to remove traces of undissolved PEO clusters. In typical experiments, droplets with initial volumes between 5μL and 50μL are left to evaporate while temperature and relative humidity are monitored. Droplets either form a disk-like solid "puddle" or a tall conical "pillar". The droplet mass is monitored using a microbalance and the droplet profile is recorded regularly using a digital camera. Subsequent processing of the data allows values of droplet volume V, surface area A, base radius R, contact angle θ and height h to be determined throughout drying. From this data we identify four stages during pillar formation: pinned drying; pseudo-dewetting; bootstrap building; solid contraction and propose physical models to explain key aspects of each stage and to predict the transition from each stage to the next. The experimental parameters of relative humidity, temperature, pressure, droplet volume and initial contact angle are all systematically varied and observed to influence the drying process and consequently whether the droplet forms a pillar or a puddle. We combine these parameters into a dimensionless Péclet number Pe, which compares the relative effects of evaporation and diffusion, and show that the drying behaviour is only dependent on $c_0$ and Pe.

## 1 Introduction

Detailed study of drying droplets is a field of research that only really began 20 years ago, with the work of Deegan et.al.[1] in which they explained the common occurrence of coffee ring stains. The final morphology of the dried solute however is not always a ring but depends on many experimental factors including: the solvent evaporation rate; interactions between solvent, vapour and substrate[2, 3]; phase transitions within the droplet[4, 5]; importance of convection currents[6]. Understanding drying processes is of great industrial importance, from the everyday applications in ink-jet printing to using nanometre sized liquid films in semiconductor devices.

The drying of millimetre sized sessile droplets is a surprisingly complex phenomena. For example, during slow evaporation in a free atmosphere, a droplet of one-component fluid evaporates with a rate that is not proportional to the surface area of the droplet, but to its radius[7]. Dust, debris and microscopic imperfections on the surface of the substrate will pin the droplet's contact line giving a fixed radius and a constant drying rate, even while the surface area decreases. A droplet of coffee is a two component system in which the water evaporates and the coffee grains do not[1]. The suspended coffee grains help to pin the contact line[8] to the substrate. Evaporation close to the contact line induces outward flow to replace the water that is lost and maintain a constant radius. This outward flow sweeps coffee grains to the edge where they are deposited, leading to the common coffee ring stain. Small droplets will dry more quickly, and the particles will not have time to migrate to the contact line and form the coffee ring stain, therefore a minimum diameter is predicted of around 10μm[9].

If concentration or temperature gradients within the liquid lead to a surface tension gradient, liquid will flow up the gradient via the Marangoni effect[10]. This is most famously recognised from 'tears of wine'. As alcohol has a lower surface tension than water, liquid in a wine glass is pulled to regions of low alcohol concentration until the liquid falls back down under its own weight. Normally flow to the contact line induces concentration gradients, and in order for a ring stain to form, the Marangoni effects must be suppressed[6].

Experiments using very concentrated colloidal silica suspensions produce different morphologies for the final solid deposit. After a period of evaporation a solid-like "gelled foot"[5] appears at the contact line, and grows towards the centre of the droplet, supporting the remaining liquid. The gelled foot is stable until further water loss and the large elastic modulus of concentrated colloidal gels causes the solid deposit to crack.[11].

In addition to single component fluids and suspensions, drying experiments have been performed on droplets of polymer solution. Polymers in solution can exist in different phases depending on the temperature and concentration. For example, highly branched polymers exhibit low glass transition temperatures[12]. The glass transition temperature is further lowered by increasing polymer concentration. This means that as concentration is increased via solvent evaporation, the concentration will eventually reach the ambient temperature glass transition concentration $c_{gt}$ and the liquid will become "glassy"[12]. Pauchard et.al observed that during evaporation of dextran (a branched polysaccharide) droplets, the concentration at the surface of the droplet increases until a glassy skin forms[4]. They propose that the skin is water permeable yet incompressible. Upon further evaporation and volume loss the skin is subjected to stress and buckles leading to various final shapes, including doughnut and sombrero-like deposits, which are predicted from initial values of contact angle, humidity, temperature and concentration[4].

The phase of a polymeric material can significantly alter its physical behaviour. Of particular relevance to this work is the "autophobic" behaviour of thin polymer films in which two molecularly identical polymers in different phases repel each other. In the case of a thin layer of liquid PDMS coated on a surface of unadsorbing PDMS, the contact line recedes with an acceleration dependant on the layer thickness and atmospheric conditions[13]. A receding contact line is also seen when droplets containing a mixture of high and low surface tension liquids are allowed to evaporate[2]. The initial droplet has a low contact angle, but if the low surface tension component is more volatile, the contact line will depin and recede when surface tension increases sufficiently. The remaining liquid, forms a smaller droplet with a much larger contact angle.

Poly(ethylene oxide), or PEO, is a very common and widely used linear polymer[14-17] and unique amongst its homologues for its unusual solubility properties[18]. It dissolves in water, although at high concentrations or molecular weights, solutions can appear cloudy due to micron-sized clusters of undissolved polymer[19]. The origin of these clusters is still a point of contention[19]. We previously studied drying droplets of PEO solution[20, 21] and showed that the solid structures that are deposited could not be described by either the ring-stain or skin buckling models, but required a 4 stage drying process: pinned drying; receding contact line; boot-strap building; late stage drying. However, in this previous work several experimental parameters were not systematically varied. Here we make significant improvements by monitoring and controlling the volume, mass, and contact angle of the droplet and the temperature, humidity and pressure in the chamber throughout the drying process. We also filtered the samples to remove undissolved clusters.

## 2 Experimental method

### 2.1 Sample preparation and characterisation

Solutions were prepared using PEO with an average molecular weight $M_w \approx 100,000$ g/mol (Sigma Aldrich 181986) and calculated radius of gyration[22] $r_g = 10$ nm giving an overlap concentration $c^* \approx 4\%$ wt. Solutions spanning a range of initial concentrations $c_0$ from 1% to 35% by mass were mixed by hand using distilled, de-ionised water and were left to equilibrate for at least 24 hours before use. A roller mixer was used to increase dissolution rate. Faster methods were not used to avoid possibility of molecular damage.

After mixing, samples appeared colourless and clear at low concentrations but above ~3%, due to undissolved polymer clusters (with average diameter around 3μm measured from microscopy) appeared cloudy. These clusters were successfully removed from all samples by driving them through a 0.45μm filter with an adjustable speed syringe pump (Harvard apparatus) at around 0.5ml per hour. Density and viscosity measurements were taken before and after filtering to monitor respectively any changes to the concentration and possible damage to the polymer from the high shear rates inside the filter. Densities were measured using an Anton Paar DMA4500 density meter giving values accurate to $0.1$ kgm$^{-3}$ and control of temperature to within $0.2°$. Viscosity was measured using Brookfield viscometer DV-II + Pro with a cone and plate geometry (Cone diameter=4.8cm and θ=0.8°) as a function of increasing and decreasing shear rate from 0 up to 900s$^{-1}$, limited to a maximum shear stress of 2.5Pa. The value of the viscosity was taken from a linear fit to the low shear data. Samples were stored in air-tight plastic centrifuge tubes until needed.

**2.2 Experimental protocol**

For each measurement, a droplet with volume in the range from 2μL to 50μL was placed onto an ethanol-cleaned borosilicate glass microscope coverslip (measuring 24mm × 50mm × 110μm from TAAB). The droplets were deposited by hand using a 1mL syringe with 0.6mm diameter syringe needle. The droplet mass was continually monitored using a Kern mass balance ALJ160-4NM to within 0.1mg and interfaced to a computer using LabVIEW. The shear rate experienced with the syringe (estimated at be $\sim 5s^{-1}$) is less than that experienced when passing the solution through the filter (estimated to be $\sim 10s^{-1}$) so we assume no damage to the polymer molecules at deposition. There may be some alignment of the polymer molecules due to the shear flow, but this has not been quantified. Atmospheric disturbances were reduced using a sealed perspex chamber (measuring 15cm × 10cm × 11cm) positioned over the droplet and the mass balance. A spirit level was used to ensure the coverslip was horizontal. Variations in temperature and relative humidity were monitored using an Omegaette HH311 probe, interfaced to the computer using the supplied software. A digital camera from ImagingSource, (model number DMK 41BU02.H) was used to record the drying process, with images taken automatically every 30 seconds using IC Capture software. The chamber was illuminated using a fluorescent StockerYale diffuse back light (ML-0405), placed behind the droplet outside the chamber. Moments after initial deposition of the droplet, the contact line becomes pinned allowing the contact angle to be altered between the advancing and receding contact angles (measured to be approximately 90° and 5° respectively) by manually adding or removing liquid with the syringe. Relative humidity in the laboratory was stable at 50±5% and was increased by introducing saturated salt solutions (sodium chloride and potassium sulphate giving 75±2% and 81±2% respectively, measured independently of predicted relative humidities[23]) or reduced to 25% by adding dried silica gel (Sigma-Aldrich) within the sealed chamber several hours before droplet deposition. This gave four values of RH between 25% and 80%. To observe the effect of ambient pressure a custom built pressure chamber fitted with a pressure gauge, digital readout and observation window was used. The pressure was reduced using a vacuum pump (Edwards E2M5) and release valve giving manual control between 20mbar and 1000mbar. To observe the effect of temperature the droplets were placed in a ceramic oven (AX series from Progen Scientific) varying temperature between 25ºC and 65ºC. The dried deposits were imaged post-experiment from the side and above.

**2.3 Data processing**

The two dimensional droplet profile (h(r)) was extracted from the digital side-on images using ImageJ software (US National Institutes of Health). We use the position of the maximum droplet height $h_{max}$=h(r=0) to define r=0 and the edge of the droplet is defined where h(r=±R)=0.

Surface area A and volume V of rotation were calculated numerically in Matlab using r=0 as the vertical axis of rotation. Uncertainties in V and A due to droplet asymmetry were quantified by calculating the difference between the contributions from the profile on either side of the rotation axis. V and A are very sensitive to variations in the position of this axis, caused by changes in the maximum point. At early times when the droplet is a spherical cap these values were compared to values of volume and area found by fitting the Young-Laplace equation to h(r), and no significant difference was seen. However once solid PEO precipitated the Young-Laplace equation could no longer be used to model the entire surface, so our numerical integration was used throughout.

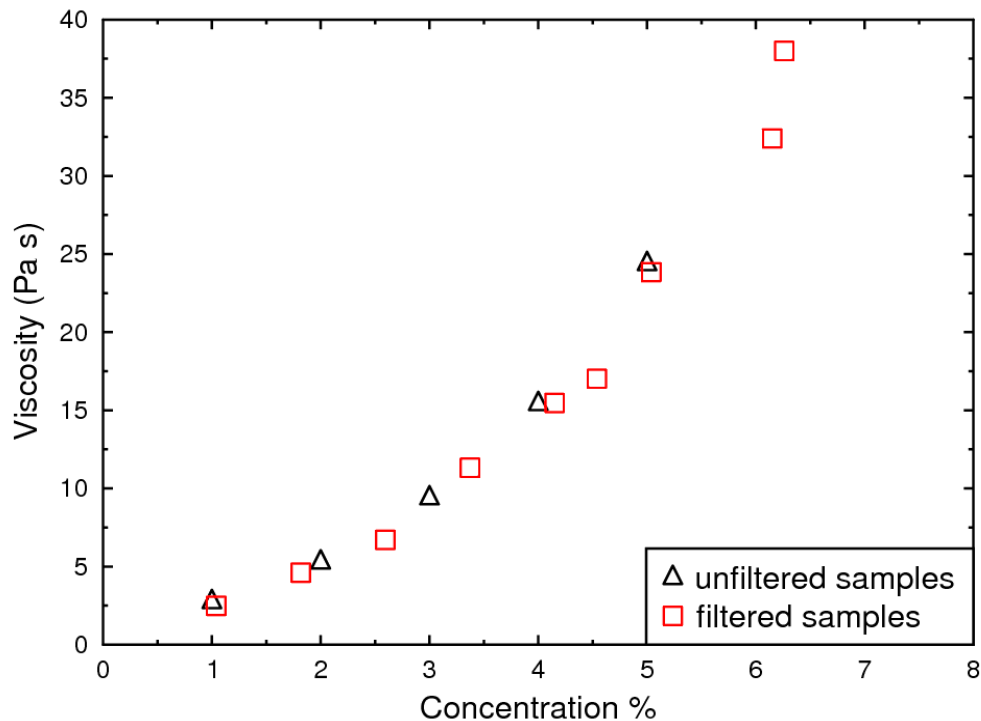

**Fig.1** Viscosity plotted as a function of concentration for filtered (squares) and unfiltered (triangles) samples. Filtering reduces concentration but does not damage polymer molecules as both sets of data lie on the same curve.

## 3 Results

### 3.1 Sample characterisation

Filtering samples removes undissolved clusters reducing the overall concentration slightly (and also density ρ). To account for this we convert filtered density values to equivalent concentrations using a linear fit through the unfiltered ρ(c) data. Fig.1 compares viscosity of filtered and unfiltered solutions at various concentrations and shows that the concentration dependence of the viscosity is unaffected by filtering. Therefore filtering is shown to reduce the density of the samples but not damage the polymer molecules as the viscosity is a sensitive measure of molecular weight.

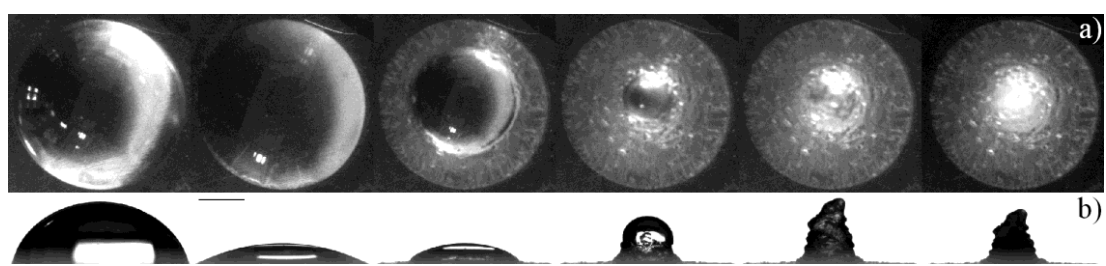

**Fig.2** Images taken simultaneously from above (a) and from the side (b) of a droplet with $c_0$=10%, $V_0$=10μL, T=22±2°C and RH=55±5%. The scale bar is 1mm and the images were taken at times 0s, 3000s, 3750s, 4110s, 4290s and 8220s respectively. Bright white patches in the side-on images are reflections of the light source. Liquid and solid phases can be clearly distinguished, with the liquid droplet being lifted by the solid in the fourth image.

### 3.2 Results from standard droplet

Fig.2 shows a sequence of images for a droplet with initial volume $V_0$=10±1μL, initial concentration $c_0$=10% at temperature T=22±2°C and relative humidity RH=55±5%. Fig.3 shows extracted measurements from this droplet over time including: volume V; surface area A; base radius of liquid droplet R, height h, and mass m, all normalised by their initial values $V_0$, $A_0$, $R_0$, $h_0$ and $m_0$ respectively. For the first 3000 seconds the droplet loses volume, height and surface area linearly, while the contact line is pinned so R remains constant. The total flux, J=-dV/dt is constant and equal to $J_0$, but the flux per unit area, j=J/A increases. This is in agreement with literature predicting higher j as θ decreases[1, 7]. As in other works[1, 20, 24], a linear fit to early volume or mass values has slope equal to $J_0$ and the intercept on the time axis defines $t_0$, the time required for the droplet to dry to zero volume at the initial drying rate,

$$t_0 = -\frac{V_0}{\left(\frac{\partial V}{\partial t}\right)_{t=0}} \quad (1)$$

Experimental times can be normalised by $t_0$ to compensate for small variations in $V_0$, RH and T.

After 3000 seconds, the rates of volume and height loss decrease, A remains constant and R starts to decrease as the contact line recedes. The time at which R begins to decrease we define as $t_1$. After another 500s, R and J continue to decrease, but h and A begin to increase. At t=4260s h and A reach maximum values $h_{max}$ and $A_{max}$, and R=0 as no liquid is visible. The deposit then contracts slowly until changes become imperceptible, but we chose to omit this late stage data from Fig.3 Error bars become significant during this stage as solid deposit becomes asymmetric.

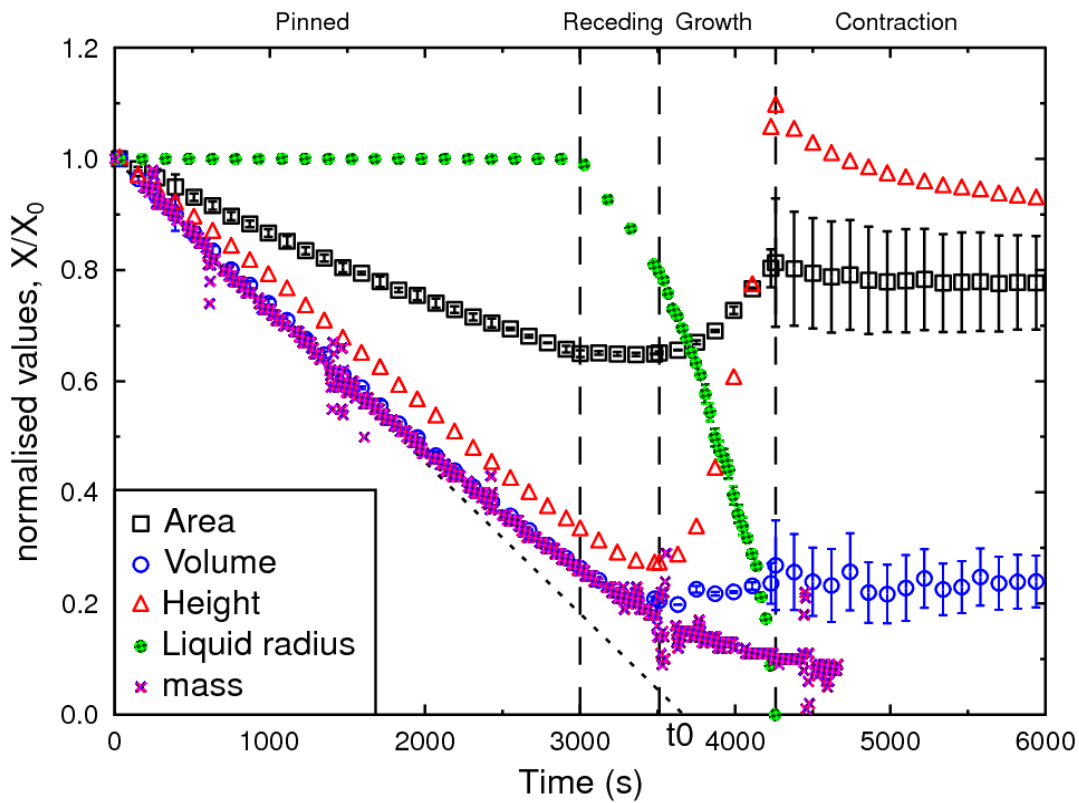

**Fig.3** Droplet volume, surface area, liquid radius, height and mass normalised by their initial values and plotted against time for the droplet shown in Fig.2. The dotted line shows the linear fit to the early values of V (or M), intercepting the time axis at $t_0$. The dashed vertical lines separate our 4 stages of drying, as labelled above the figure and described in the text.

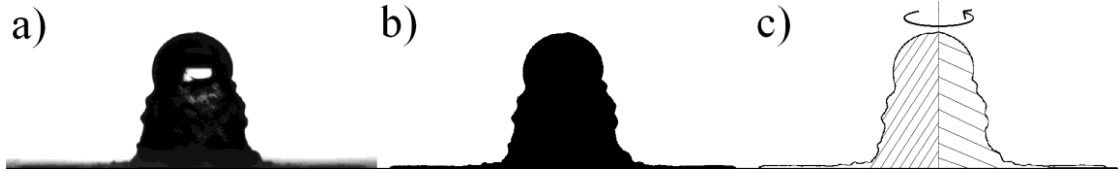

**Fig.4** shows the method of calculating surface area and volume from the droplet profile image a). Using ImageJ software the image is converted into binary b) and a 2d surface profile is extracted c). Surface area A and volume V were integrated numerically using the position of $h_{max}$ as the axis of rotation. We account for asymmetry in the profile by calculating the volume and surface area given by the half profiles to the left and right of the axis. We add the left and right values to obtain the mean, and use the difference for the uncertainty.

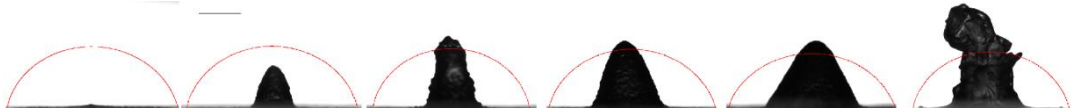

**Fig.5** shows final deposits of droplets with $c_0$=3, 5, 10, 15, 20 and 25%. The scale bar is 1mm long.

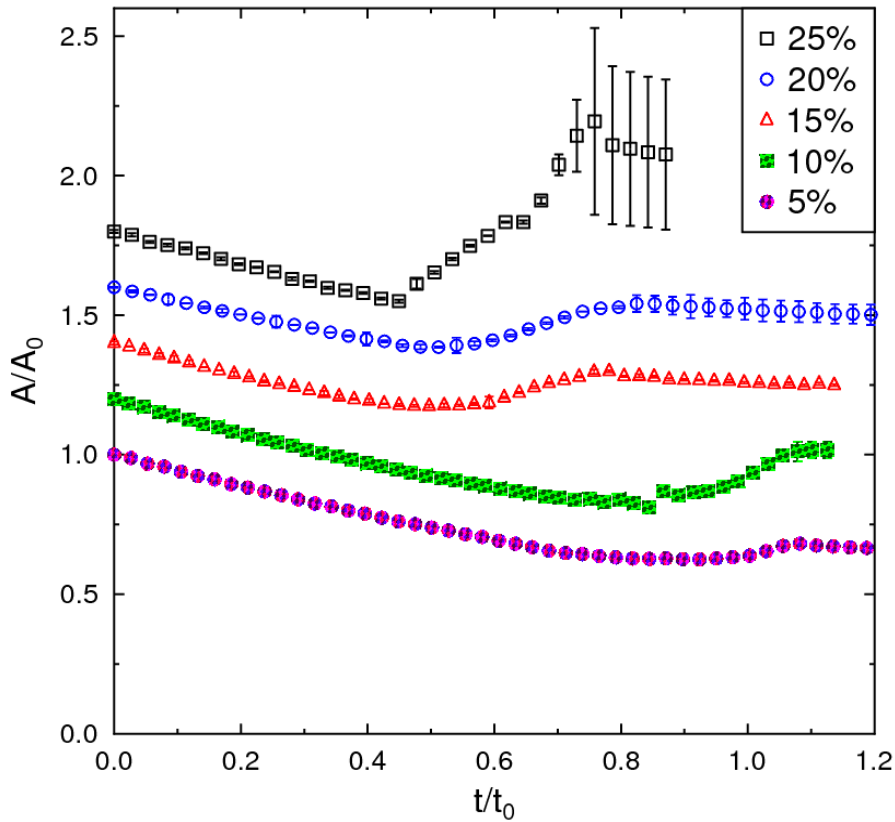

**Fig.6** shows normalised area $A/A_0$ against normalised time $t/t_0$ for droplets with $V_0$=10±1µl, $\theta_0$=70±5°, RH=55±5%, T=22±2°C and $c_0$=5, 10, 15, 20 & 25%. We observe an increase in surface area during the vertical growth stage for $c_0 \geq 5$%. When $c_0 < 5$% pillars do not form and no surface area increase is observed.

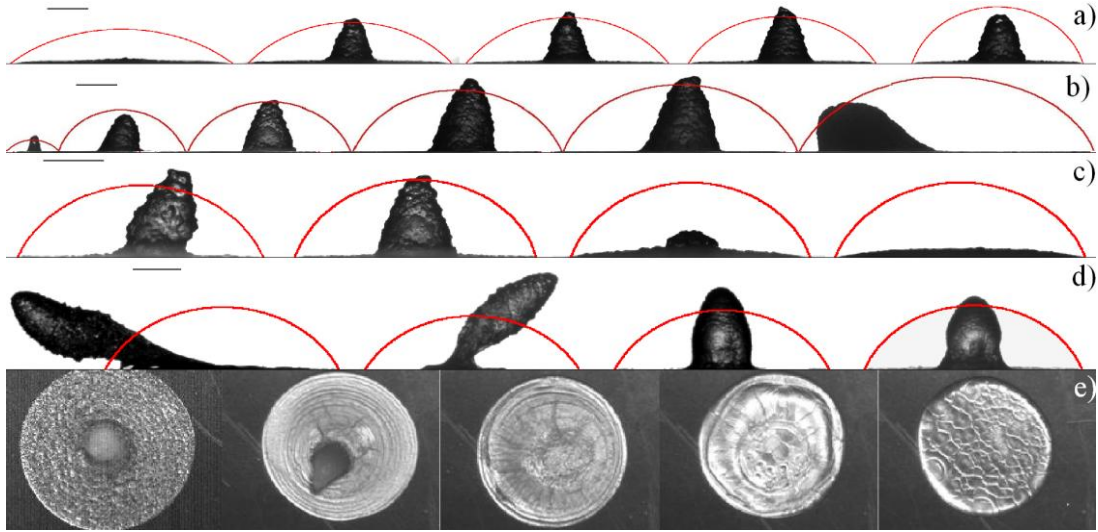

**Fig.7** is a composite image showing samples with varying (a) contact angle, $\theta_0$=40, 47, 54, 57 and 70° (b) initial droplet volume, $V_0$=0.4, 5, 10, 20 30 and 50μL (c) relative humidity, RH=25, 55, 75, 80% (d) pressure, P=20, 50, 100, 200mbar and (e) temperature, T=22, 30, 40, 50 and 60°C, for droplets with $c_0$=10% and unless specifically varied $V_0$=10±1μl, $\theta_0$=70±5°, RH=55±5% and T=22±2°C. The thin curves (red on-line) show the initial droplet profile. The scale bar is 1mm long for each respective figure.

### 3.3 Effects of varying parameters

We systematically alter the initial droplet parameters ($c_0$, $\theta$ and $V_0$) and atmospheric conditions (RH, T, and pressure P) and describe the effects below.

**Varying concentration** In Fig.5 we compare the final deposits of 5 droplets with $c_0$=5, 10, 15, 20 and 25%, $V_0$=10±1μl, $\theta_0$=70±5°, RH=55±5% and T=22±2°C. The size of the final deposit increases with concentration.

In Fig.6 we compare the evolution of $A/A_0$ for five representative values of $c_0$ with time normalised by $t_0$. We see a period of linear decrease in surface area during stages 1 & 2, followed by a period in which even within our realistic error bars, all values of $c_0 \geq 5\%$ show a significant increase in surface area during stage 3. When $c_0 < 5\%$ no surface area increase occurs so this data was omitted from the graph.

Fig.7 is a composite image showing samples with varying (a) contact angle (b) initial droplet volume (c) relative humidity (d) pressure and (e) temperature, for droplets with $c_0$=10% and unless specifically varied $V_0$=10±1μl, $\theta_0$=70±5°, RH=55±5% and T=22±2°C.

**Contact angle** Figure 7a compares final deposits for droplets with $\theta_0$=40, 47, 54, 57 and 70º and shows that there is a lower limit for $\theta_0$ below which pillar formation is supressed. For these droplets this limit is between 40º and 47º.

**Volume** Figure 7b compares final deposits with $V_0$=0.4, 5, 10, 20 30 and 50µL. For $V_0 \leq 30$µL the height of the deposit increases with $V_0$ and is proprtional to $h_0$. Above 30µL, where the initial droplet is no longer a spherical cap, the deposit collapses during the growth stage. Our droplet deposition method limits the minimum value of $V_0$. This minimum has a much larger diameter than that required for polymer migration to the contact line to be supressed, D~10µm[9].

**Relative Humidity** Figure 7c compares final deposits with RH=25, 55, 75 and 80% and shows that there is an upper limit for RH above which pillar formation is supressed. For these droplets this limit is between 75% and 80%. For RH$\geq$90% drying is too slow for practical experiments. In a saturated environment (RH=100%) dried deposits absorb moisture and revert to liquid droplets.

**Pressure** Figure 7d compares final deposits for droplets in pressures of 20, 50, 100 and 200mbar. For these measuremnets data is more qualitative due to restrictuions of the pressure chamber and relative humidity and temperature were neither controlled nor measured. The drying was significantly faster, with the 20mbar droplets drying in around 15 minutes. When P$\leq$50mbar we observe bubbles forming within the droplet which are dissolved gases coming out of solution and disrupt early drying (at room temperature water boils between 20 and 30mbar[25]). However the same 4 stages were still observed, although with noticable differences: $t_1/t_0$ was very low (~0.2 compared to typical values of between 0.65 and 0.75 for $c_0$=10% at 1 atmosphere); at the end of stage 2 $\theta$>>90º and R is very small; the solid structures that form are slender and more unstable so do not always grow vertically.

**Temperature** Figure 7e compares the final deposits for droplets in 22, 30, 40, 50 and 60ºC. Due to size of the oven it was impossible to record side profiles so we only have images from above and again relative humidity was neither controlled nor measured. We observe an upper limit between 30ºC and 40ºC above which pillars do not form. Instead at 40ºC and 50ºC the final solid deposit is a smooth flat disk. At 60ºC cracks appear in the disc. In a separate experiment the melting temperature of solid PEO was found to be between 65ºC and 70ºC.

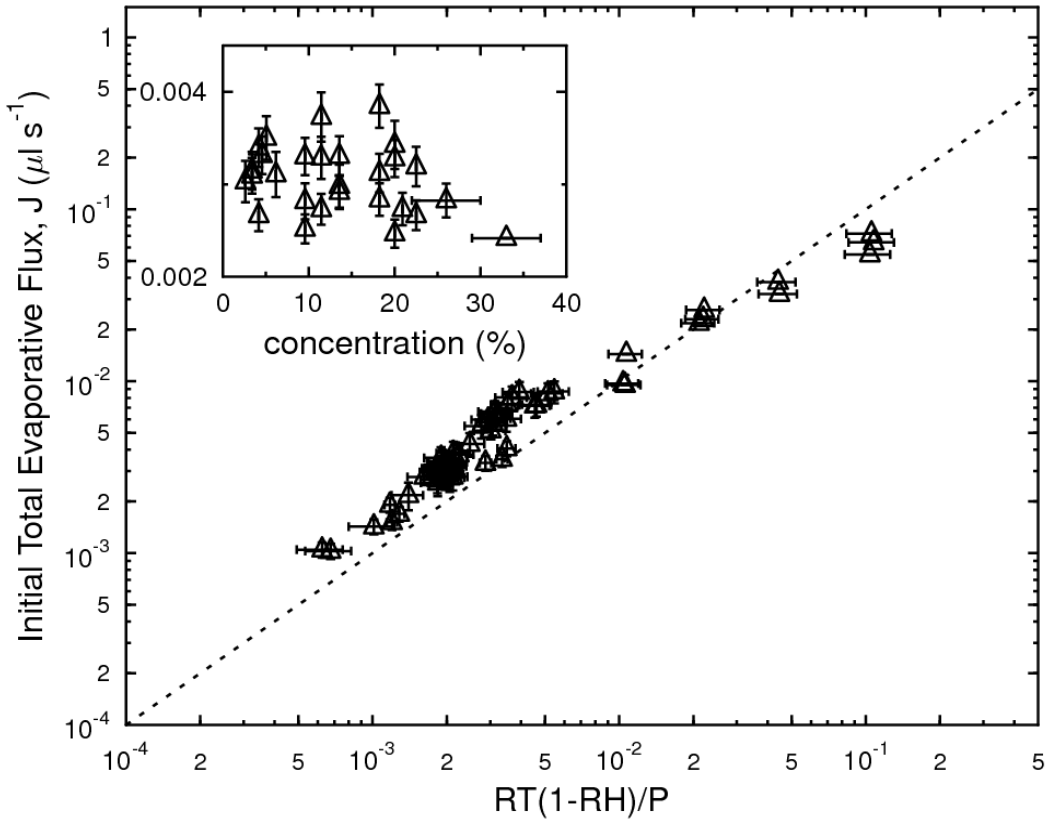

**Fig.8** A log-log plot of total evaporative flux J against R*T*(1-RH)/P showing power law behaviour with slope of 1. In the insert, J is plotted against $c_0$ for constant $R_0$ and shows no systematic variation with initial concentration.

## 4 Discussion

### 4.1 Four-stage model of standard droplets

As discussed in our previous work, and supported by data presented here (in particular Fig.6 showing increasing surface area) the drying process observed here is different to the skin-buckling effect observed in other drying polymer solution systems[4, 24]. The drying timeline, as shown in Fig.3, is split into 4 stages, detailed below:

**Stage 1** During Stage 1 the droplet shows typical pinned contact line drying behaviour with a constant droplet radius R. To accommodate the reducing droplet volume, h and $\theta$ both decrease but typically $\theta$ remains above the receding contact angle (measured in separate experiments to be around 5° for $c_0$=15%). The evaporation rate is greatest at the contact line (provided $\theta<90°$), and to remain pinned must be sustained by solvent within the droplet flowing radially outwards[1].

In agreement with other work [6], Fig.8 shows that $J_0 \sim R_0\, T\, (1-RH)/P$, so the flux is dependent on droplet size, temperature, humidity and pressure, but independent of contact angle or concentration. This is to be expected for droplets of pure liquid as the lower surface area for a low contact angle droplet is countered by the increasing flux at the contact line[24]. Perhaps more surprising is the observation that the flux is independent of polymer concentration (as shown in Fig.8 inset) – higher polymer concentration at the surface means lower water concentration and therefore a lower evaporation rate would be expected. However this observation can be explained as PEO has been shown to reduce surface tension at the water-air interface by forming a molecular layer there. The amphiphilic properties of PEO mean the hydrophobic $CH_2$ units will preferentially go to the interface and the O units will hydrogen bond with the water. Once the surface is fully covered by PEO (at concentrations above ~0.01%), it has been shown[26, 27] that while an increase in $c_0$ in the bulk of the liquid droplet will result in more PEO molecules at the interface, their arrangement is such that the number of $CH_2$ groups remains constant, so the water concentration is also constant. Therefore J is independent of $c_0$ and $\theta_0$.

**Stage 1 to stage 2 transition** When the droplet concentration at the contact line reaches the saturation concentration, $c_{sat}$, solid PEO precipitates as solid semi-crystalline spherulites. It is assumed that once spherulites have been deposited any remaining water is trapped and cannot diffuse back into solution. These spherulites cause evaporation through the surface at the contact line to stop, which in turn stops radial flow. Without radial flow the liquid cannot remained pinned and so we observe a receding contact line as soon as precipitation begins.

The time at which $c_{sat}$ is reached, $t_1$, depends on $c_0$ and how quickly polymer chains build up at the contact line. By increasing the evaporative flux and keeping the co-operative diffusion coefficient constant, the polymers motion to the contact line will be faster and therefore $t_1$ is reduced.

Using Fick's law, the local evaporative flux can be written as $j=D_c\nabla c$, where $D_c$ is the cooperative diffusion coefficient and $\nabla c$ is the concentration gradient, given by the concentration difference between the edge (equal to $c_{sat}$ at $t_1$) and the droplet centre (assumed to still be $c_0$ at $t_1$) divided by a diffusive length scale, on the order of, $(c_{sat}-c_0)(D_c t_1)^{-1/2}$. Combining gives $j t_1^{1/2} = D_c^{1/2}(c_{sat}-c_0)$. In Fig.9 we plot $j t_1^{1/2}$ against $c_0$, with the best fit line crossing the horizontal axis at $c_{sat}=60\pm6\%$ in close agreement with literature values[22]. From the gradient we calculate a reasonable value for the diffusion coefficient as $D_c=2.5\pm1\times10^{-10}$ cm$^2$s$^{-1}$. More exact 1D models (for example incorporating moving boundary conditions[28] lead to predictions which do not fit the data as well. A full 3D calculation, including fixed boundary conditions around the contact line but moving interface over the droplet surface, coupled to convective terms driving the polymer towards the edge are needed to accurately model the data, but are beyond the primarily experimental focus of this work.

**Stage 2** As the contact line recedes solid crystallites are continually preciptated at the contact line and deposited as a thin solid layer. Due to competition between the receding contact line increasing h and the evaporation reducing h, there is often a period in which h continues to decrease, but more slowly than in stage 1. We call this period stage 2 which ends when h begins to increase.

The receding of the liquid phase can be explained by examining the unbalanced surface forces acting at the contact line, and dissipating this force through a receding wedge of liquid[29] – preliminary measurements suggest that the speed of the interface is around 1μm/s, consistent with the velocity calculated using receding velocity $v=\theta^3\gamma/\eta$, where surface tension $\gamma=50$mN/m[28], viscosity $\eta=50$Pa.s[30] and contact angle $\theta=5°$. However, the receding is always preceded by solid deposition, and no specific contact angle at $t_1$ is required. This suggests that surface tension is not the driving force behind stage 2 receding. Alternatively, the growth of the crystalline layer could be driving the receding liquid.

**Stage 3** During this stage h and A begin to increase and θ continues to increase. Around θ=90° solid spherulites begin to deposit on top of previous deposits, lifting up the edge of the liquid droplet. Due to continuing evaporation, the droplet radius continues to reduce, resulting in a solid conical structure. As with stage 2, this stage is also driven by solid precipitation, but they are differentiated by whether the droplet height is decreasing (stage 2) or increasing (stage 3).

In cases where $c_0$ is high (>25%), stages 1&2 are very short, and stage 3 is longer, resulting in very large and often unstable pillar formations. Stage 3 ends when the structure reaches its maximum height and the outer layer has solidified; liquid may still be present within.

Some droplets, typically those with lower intial concentrations, never reach Stage 3: h steadily decreases and the final deposit is a flat rough puddle, with similar thickness everywhere. Mathematically this can be seen by investigating the height change of a spherical-cap droplet as the contact line recedes and the volume reduces through evaporation. A minimum height at $t_1$ is predicted (proportional to the evaporation rate divided by receding velocity) below which h always decreases and a pillar will not form.

**Stage 4** When the remaining liquid is completely encapsulated by solid PEO the drying rate reduces significantly. Further volume loss leads to the solid structure formed at the end of Stage 3 slowly shrinking until the deposit is completely dry. Because this process is so slow no definitive duration is measurable, however the opacity of the deposit is a good indication of whether it is completely dry or not. During this stage, the forces generated by the shrinking structure stuck to the coverslip can be sufficiently strong to cause the glass coverslip to bend upwards[31].

**4.2 Unifying behaviour using Péclet number**

The effects of initial contact angle, volume, relative humidity, temperature and pressure on whether a pillar will form or not can all be understood in terms of the droplet's initial Péclet number. The Péclet number is a ratio between the effects of flow of polymer due to the evaporative flux per unit area, j, and the co-operative diffusion coefficient $D_c$ of the polymer, made dimensionless by multiplying by the smallest dimension of the droplet, which for all of our droplets with $\theta_0 < 90°$ is the initial height $h_0$:

$$Pe = \frac{h_0 j}{D_c} \qquad (2)$$

In order for precipitation to take place preferentially at the contact line, the concentration here must be higher than elsewhere. The local evaporative flux from the pinned contact line during stage 1 induces outward flow of the polymers, as described in literature[1]. Opposing this motion is the polymers co-operative diffusion, which tends to drive the polymer to homogeneity. In order for deposition at the contact line to occur, the effect of the evporative flux must be greater than the effect of diffusion. The Péclet number encapsulates this competition: a low Péclet number would lead to shallow concentration gradients, no preferential deposition at the edge and a puddle-like final deposit; a high Péclet number would give very early crystallisation at the contact line, followed by a receding contact line and increasing height during stage 2, and a final pillar-shaped deposit.

Below we explain the results of all the varying conditions using the Péclet number model. Values of $j_0$ and $h_0$ were taken from profile and gravimetric analyses. Unless specified otherwise $D_c$ is assumed to be constant at $2.5\pm1\times10^{-10}$ cm$^2$s$^{-1}$, as measured from Fig.9, and concentration is at $c_0$=10%.

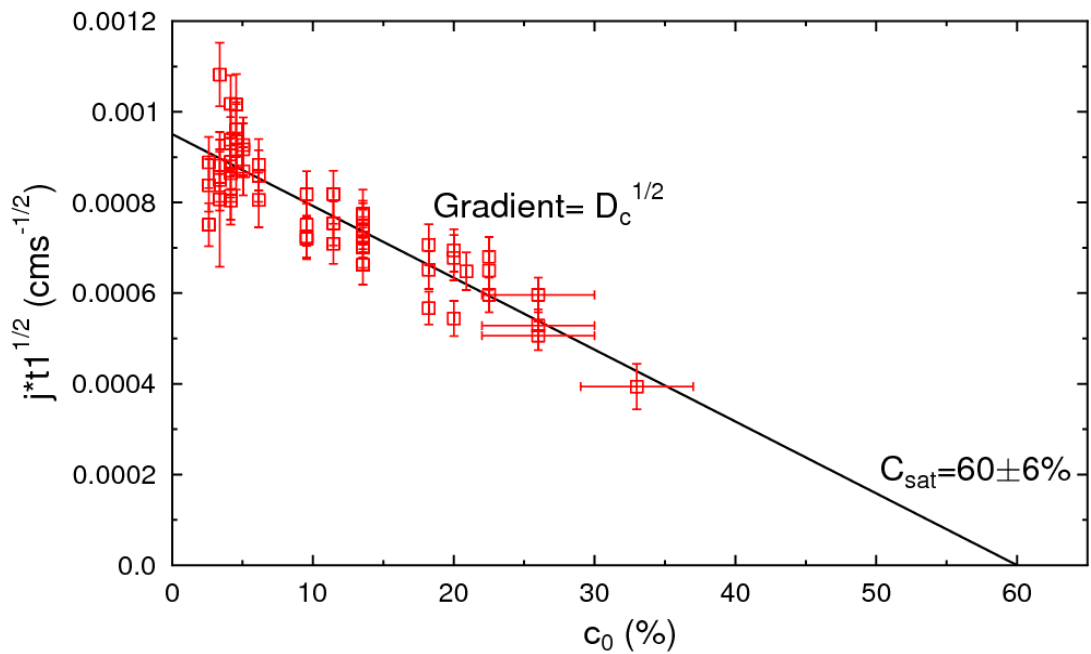

**Fig.9** shows $jt_1^{1/2}$ plotted against concentration and shows a good straight line fit. This gives $c_{sat}$=60±6% in close agreement with literature [19] and $D_c$=2.5±1×10$^{-10}$ cm$^2$s$^{-1}$.

**Relative Humidity and Pressure** At constant concentration and temperature, $D_c$ is constant. Increasing j by reducing pressure or relative humidity will increase Pe. Pillars continue to form, and under low pressures are significantly taller than at atmospheric pressure. Conversely, reducing j by increasing RH stops pillar formation, as expected for lower values of Pe.

**Volume** For a spherical-cap droplet at constant contact angle, $h_0$~R and $A_0$~R$^2$. We have also shown (Fig.7) that J~R so j~R$^{-1}$. Provided that the droplet is spherical, we can say that Pe~R$^0$ so will not be affected by the droplet volume, as shown by the similarity of the final deposit shapes when $V_0\leq30\mu L$ in Fig.6b. However, at higher volumes the capillary length of the liquid (~2mm for water)[29], causes the droplet to lose its spherical cap shape so this argument no longer holds, and may explain the anomalous collapsed pillar structure observed for $V_0=50\mu L$.

**Contact Angle** It is known that J∝R, so by altering the contact angle at constant volume the Péclet number is altered as the radius, height and surface area are all affected. Using $A_0=\pi R^2(1+X^2)$, where $X=h/R=\tan(\theta/2)$ we can write;

$$Pe = \frac{X}{1+X^2} \quad (3)$$

giving a maximum at h=R and θ=90° and a minimum Pe=0 at θ=0° in agreement with experiments showing a cut-off θ below which pillars do not form. The equation also predicts Péclet number will be reduced when θ>90°, however this was untestable without altering the substrate chemistry.

**Temperature** From literature it is known that in the temperature range we are working at, total evaporative flux is linearly proportional to temperature $J \sim T^1$[11]. When $c_0>c^*$ (the overlap concentration) the cooperative diffusion coefficient is a function of concentration, temperature and viscosity[33] $D_c=(1-c_0/100)^2 kT/(6\pi R_H \eta)$ where k is Boltzmann's constant, $R_H$ is the apparent hydrodynamic radius and η is the viscosity. Previous work studying the viscosity of PEO $M_w=10,000$g/mol[34] shows that in the range of temperatures we have observed $\eta \sim T^{-\alpha}$ where α lies between 2 and 3. Combining these various dependencies on T, at constant concentration, gives;

$$Pe \sim T^{-\alpha} \quad (4)$$

Thus an increase in temperature will lead to smaller Peclet number, as the effects of faster evaporation are insignificant compared to the reduction in viscosity and increase in diffusion. At higher temperatures we observed reduced pillar formation, consistent with lower values of Pe.

**Combining $c_0$ and Pe** In Fig.10 we plot the Pe-$c_0$ phase diagram of droplet behaviour marking the droplets that pillar with a triangle and droplets that puddle with a hollow circle. By considering the effects of V, θ, RH, P and T in terms of the Péclet number, the behaviour becomes universal. For example of the 4 droplets with $c_0=10$% that form puddles, 2 had low θ and 2 had high RH but all had similar values of Pe and behaved similarly. The low pressure droplets have very high values of Pe and are off the vertical scale, but are still consistent. We do not currently have a prediction of where the transition should be, however it is clear that a well defined boundary separating pillars from puddles exists. From Fig.8 we would expect that low concentration droplets would form pillars if the Péclet number is sufficiently high, e.g. in low pressures.

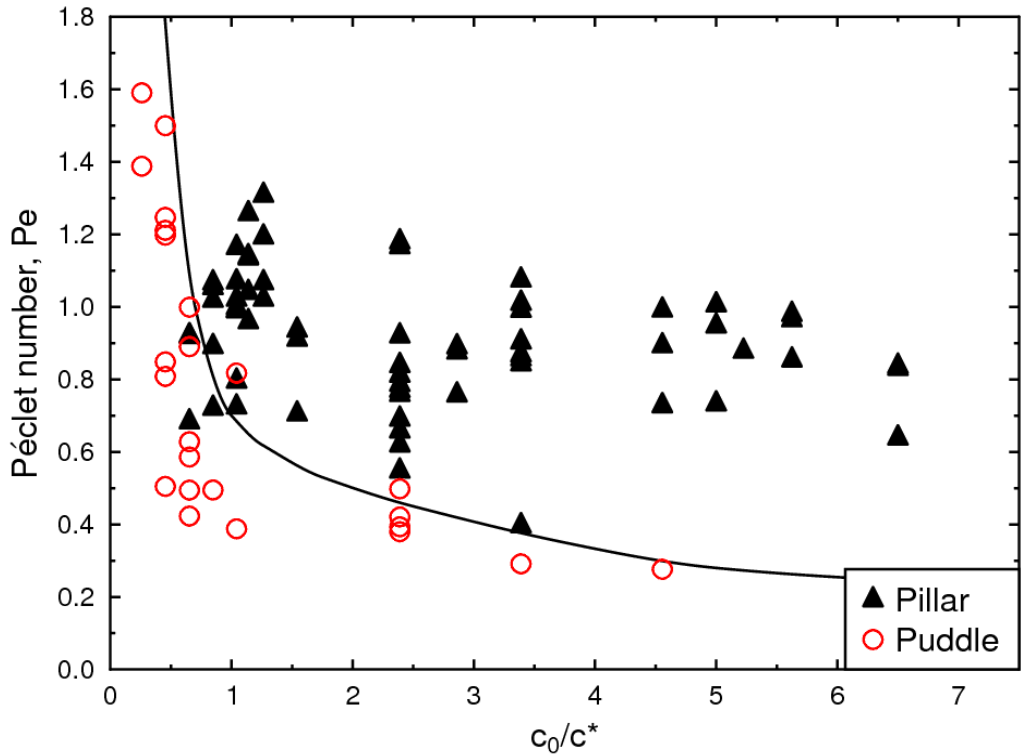

**Fig.10** Péclet number Pe is plotted against concentration $c_0$, with filled triangles representing pillars and hollow circles representing puddles. The behaviour of all samples is dependent only on these two parameters, allowing predictions to be made: we expect low concentration droplets with high Pe to also form pillars. The line is a guide to the eye.

## 5 Conclusions

Sessile droplets of aqueous poly(ethylene oxide) solution dry via an unusual 4 stage process: following typical pinned drying (stage 1) in which droplet height and volume decrease and polymer concentration increases at the edge of the dropet due to outwards capillary flow, in stage 2, initiated by precipitation of solid material, the contact line depins and begins to recede while the height continues to decrease. For many droplets, the height then increases (stage 3) as the edge of the liquid droplet is lifted up by the solid deposit, sometimes to over twice the initial height. A solid conical pillar forms, which during Stage 4, continutes to dry and shrink.

The final shape of the deposit is shown to depend on many experimentally controllable parameters:

- At low concentration, droplets dry to flat puddles, whereas at higher $c_0$ pillars become taller and unstable. The height of the droplet at the end of stage 1 is critical in determining which behaviour occurs.
- At high humidity, all concentrations form puddles.
- At low pressure, very tall unstable pillars are observed.
- At higher temperatures, pillars become less pronounced.
- Droplets with low initial contact angle do not form pillars.
- Initial droplet volume does not significantly alter deposit shape, provided the droplet is a spherical cap.

The effects of all these parameters except concentration are combined using the Péclet number, a dimensionless ratio comparing the relative effects of flux and diffusion, and a phase diagram of Pe-$c_0$ shows universal behaviour. Further work is now required to test additional predictions and to theoretically predict the boundary between pillar and puddle formation.